# Supercurrents in Magnesium Diboride/Metal Composite Wire

M. J. Holcomb

Nov̄é Technologies, Inc.
3380 South Lapeer Road
Metamora, Michigan 48455 USA

(main) 810-678-8014 (fax) 310-388-1370 (email) mholcomb@novewire.com

**Abstract**

We have fabricated a series of *ex situ* copper sheathed powder-in-tube $MgB_2$ wires with 20% by volume Ag, Pb, In, and Ga metal added to the $MgB_2$ powder. We find the transport critical current of these wires increases significantly with the addition of specific metals to the core filament. In particular, the critical current density ($J_C$) of the $MgB_2$/Ga(20%) wire is in excess of $5 \times 10^4$ A/cm$^2$ at 10K in self field, nearly 50 times that of the $MgB_2$/Ag(20%) wire. The temperature dependent $J_C$ of all wires is well described as an ensemble of clean *S/N/S* junctions in which the relevant parameters are the average thickness of the *N* layer, the critical temperature of the *S* layer, and a scaling term related to $J_C$ at zero temperature. Eliminating the differences in the filament microstructure as the primary cause of the enhanced $J_C$, we suggest that $J_C$ is determined by the magnitude of the proximity effect induced superconductivity in the normal metal layer, which is known to be proportional to the electron-electron interaction in *N*. We present one-dimensional material specific calculations that support this, and zero-field cooled DC magnetic susceptibility data that confirm an increased number of well-connected superconducting grains exist in the composite wires that contain metal additions with large electron-electron interactions and long electron mean free paths.

## I. Introduction

The discovery of superconductivity near 40K in magnesium diboride ($MgB_2$) [1] sparked a worldwide effort to develop technically useful $MgB_2$-based conductors for use in superconducting applications at temperatures less than 30K. Recent improvements in the current carrying properties of $MgB_2$ wire and tape in modest magnetic fields [2, 3], and the low materials cost of $MgB_2$, suggest that these conductors may be economically competitive with technical NbTi and $Nb_3Sn$ wire at 4.2K. Further, for superconducting applications in the 20K to 30K temperature range, which precludes the use of both NbTi and $Nb_3Sn$ wire, $MgB_2$ wire and tape represent an inexpensive alternative to commercial high-temperature superconductor (HTS) tapes. With a significantly lower materials cost relative to HTS tape, the development of high critical current $MgB_2$ wire and tape may enable the economic adoption of superconducting devices such as transformers, motors, generators, and MRI magnets operating at temperatures above 20K in modest magnetic fields.

Similar to HTS ceramics [4], $MgB_2$ is mechanically brittle and not easily drawn into a fine wire. Unlike HTS ceramics, however, supercurrent flow in polycrystalline $MgB_2$ is not significantly impeded by the grain boundaries due to the relatively long coherence length of the material [5, 6] and the metallic nature of the grain boundaries [7]. These characteristics have allowed for the relatively rapid development of $MgB_2$ wire and tape with critical current densities ($J_C$) in the range of $10^4$ to $10^6$ A/cm$^2$ at 4.2K in modest magnetic fields ( < 4 Tesla ) by a number of groups worldwide [8, 9, 10, 11, 12, 13, 14, 15, 16, 17, 18, 19, 20, 21, 22, 23, 24].

Flükiger has recently reviewed the development status of MgB$_2$ wires and tapes [25]. Although a number of methods have been investigated to fabricate MgB$_2$-based conductors, the most successful to date is the powder-in-tube (PIT) method. In general, MgB$_2$ wire and tape fabrication is achieved using so-called *in situ* or *ex situ* PIT methodologies. In the *in situ* PIT method, stoichiometric amounts of magnesium and boron powder are well mixed and packed into a metal tube, which is sealed and drawn into wire. The wire is typically annealed between 600°C to 950°C with a variety of heating schedules to form a well-connected superconducting MgB$_2$ central filament. Critical current densities of *in situ* MgB$_2$ PIT wires are limited by filament porosity, partially reacted precursor material, and deleterious reactions of Mg, B, or MgB$_2$ with the metal sheath material. A distinct advantage of the *in situ* PIT method is the ability to introduce nano-sized artificial pinning centers to the MgB$_2$ core filament [2], which has been shown to significantly increase the $J_C$ of these conductors in applied magnetic fields. In contrast, the *ex situ* PIT method of fabricating MgB$_2$ wire and tape is very similar to the *in situ* method except for the fact that MgB$_2$ powder is packed into the metal billet. Although critical current densities in excess of $10^5$ A/cm$^2$ at 4.2K in self-field [12] have been reported on unsintered *ex situ* $MgB_2$ tape, a final high temperature anneal, which establishes a well-connected superconducting filament, generally leads to a significantly increased $J_C$ in these conductors. A number of factors have been identified which affect the final $J_C$ of *ex situ* $MgB_2$ conductors, these include the phase purity and morphology of the $MgB_2$ powder, the reactivity of $MgB_2$ with the metal sheath, and the final heat treatment schedule. Despite a number of factors which limit the $J_C$ of *in situ* or *ex situ* $MgB_2$ wires and tapes, development has proceeded steadily and transport $J_C$s in excess of $10^5$ A/cm$^2$ at 4.2K in self-field have been achieved in small coils made using multiple meter lengths of $MgB_2$ PIT wire and tape [13, 16, 26].

In addition to possessing high $J_C$s in applied magnetic fields at temperatures less than 30K, technical $MgB_2$ wire and tape must also possess good thermal and mechanical stabilization. This depends critically on the metal sheath material used to encase the $MgB_2$ filaments. It is well known that high current NbTi and $Nb_3Sn$ wire achieve stability by fabricating the wire with many thin superconducting filaments, which are embedded within a high purity stabilizing matrix [27] such as copper or aluminum. In order to fabricate stabilized multi-filamentary $MgB_2$ conductors, the sheath material must also possess sufficient ductility and strength to lend itself to common PIT wire and tape fabrication techniques, and must be relatively unreactive to Mg, B, or $MgB_2$ at the required annealing temperatures. A variety of sheath materials for $MgB_2$ wires and tapes have been explored in the literature including Cu, Ag, Fe, Ni, Nb, Monel, and stainless steel [8-26]. Although copper is a very good, low cost stabilizing material, the tendency for Cu to form intermetallics with Mg at temperatures in excess of 650°C makes it less satisfactory as a sheath for $MgB_2$ [28, 29, 30].

The vast majority of high $J_C$ $MgB_2$ wires and tapes in the literature are sheathed in either Fe or Ni. Although less reactive than Cu, these materials have also been found to react with Mg, B, or $MgB_2$ at temperatures in excess of 850°C forming $Fe_2B$ [31] or $Mg_2Ni$ [32], respectively. Further, the use of Fe, or Ni as sheath materials, with high electrical resistivities and low thermal conductivities with respect to high purity copper, results in conductors which tend to quench at transport currents well below the intrinsic $I_C$ of the $MgB_2$ filament. This is implicit in the literature data on Fe- or Ni-sheathed $MgB_2$ wire and tape in that most high $J_C$s ($>10^4$ A/cm$^2$) are obtained inductively, and many high transport $I_C$s are obtained with short duty-cycle current pulses. Overall, the development of high $I_C$ $MgB_2$ wire and tape with adequate thermal and mechanical stabilization remains a significant barrier to the adoption of these conductors in superconducting applications at temperature below 30K.

In principle, the fabrication of high $I_C$ copper sheathed *ex situ* $MgB_2$ PIT wire is also feasible if it is possible to establish well-connected $MgB_2$ grains *without* employing high-temperature heat treatments. For example, increased $I_C$s in unsintered *ex situ* $MgB_2$ PIT wire have been observed by a number of groups with the addition of a normal metal powder to the MgB$_2$ core filament [33, 34, 35, 36]. In particular, Tachikawa *et al.* [33, 34] reported a J$_C$ enhancement of nearly a factor of five in MgB$_2$ PIT wire with the addition of 10% by volume indium to the core filament and employing a low temperature (< 200°C) heat treatment. While this increase is much less than that realized through high temperature anneals of *in situ* or *ex situ* MgB$_2$ wire, it does suggest the use of normal metal additions to as a means of increasing the density of well-connected MgB$_2$ grains in the filament. It has been proposed that the increase of J$_C$ in MgB$_2$ PIT wire with metal additions to the core filament may be due to an increase of the intra-filament thermal stability resulting from the addition of a metal with low electrical resistivity and high thermal conductivity [36], an improvement in the MgB$_2$ "grain linkage" presumably resulting from an increased Josephson current between the grains, or a proximity effect within the normal metal [34], an increased densification of the core filament, or a combination of all of these factors.

Here we report the systematic study of the effects of normal metal additions (Ag, In, Pb, and Ga) to *ex situ* copper sheathed MgB$_2$ PIT wire fabricated with no post deformation heat treatments. We find the core density and microstructure of the wires with 20% by volume Ag, In, and Pb to be nearly identical. However, the I$_C$ (10K, self-field) of the wire with 20% by volume In metal added to the filament is over 30 times greater than that of the wire with 20% by volume Ag at 10K. We also find that, unlike the wires with Ag, In, and Pb added to the filament, the filament of the $MgB_2$ PIT wire with 20% by volume Ga is inhomogeneous on the length scale of the filament diameter and displays significant alloying with the sheath at the Cu/filament interface. Despite this, we find that the $I_C$ of the $MgB_2$ PIT wire with 20% by volume Ga is nearly 50 times greater than that of the $MgB_2$ PIT wire with 20% by volume Ag, and over 60 times greater than that of the $MgB_2$ PIT wire with no metal addition. We suggest that the primary reason for the significantly increased $I_C$ of these wires is an increased $MgB_2$ grain connectivity due to the superconducting proximity effect in the normal metal, and not simply the result of increased filament density or intra-filament thermal stability.

This paper is divided into five sections. In section II we describe briefly the fabrication and characterization of all $MgB_2$-based PIT wires reported in this study. In section III we describe the theoretical model used to interpret these data, and then discuss the implications of these results in section IV. In the Conclusions section, we summarize our results and discuss further this approach to the fabrication of high $I_C$ *ex situ* $MgB_2$ conductors, with emphasis on the technical requirements necessary for superconducting device applications at temperatures less than 30K.

**II. Experimental**

A series of copper sheathed *ex situ* $MgB_2$ monofilament wires has been fabricated using standard techniques [4]. All wires were made with 20% by volume metal added to the core filament. Previous studies indicate that the maximum $J_C$ at 4.2K in self-field is obtained with approximately 15% to 35% by volume metal addition for both $Nb_3Sn$/Metal [37] and $MgB_2$/Metal [38] PIT wires using -325 mesh superconductor powders. These data and similar results obtained by Tachikawa *et al.* [34] suggested the selection of 20% by volume metal addition for all wires in this study.

The $MgB_2$ powder used to fabricate all samples was obtained from Alfa Aesar (-325 mesh, 99.9% metals basis). The as-received $MgB_2$ was sieved in an argon inert atmosphere box using a 450 mesh (32μm) stainless steel sieve. Only the -450 mesh fraction of the $MgB_2$ was used in the fabrication of the wires in this study. In addition to a control $MgB_2$ PIT wire, four additional wires were fabricated using 20% by volume metal added to the $MgB_2$ powder [39]. The metal additions consisted of silver (0.5-1 μm, 99.9%, Alfa Aesar), indium (-325 mesh, 99.99%, Alfa Aesar), lead (-200 mesh, 99.9%, Alfa Aesar), and gallium (liquid, 99.999%, Alfa Aesar).

All $MgB_2$/Metal(20% by volume), $MgB_2$/M(20), composites were prepared using a Fritsch planetary mill housed in an inert atmosphere box under argon (< 1ppm $O_2$). The milling of the composites was designed to homogenize the component materials and not necessarily to reduce the $MgB_2$ particle size. The materials were milled at 500RPM for 2 hours in an 80ml stainless steel vial containing five 10mm diameter WC balls and one 25mm diameter WC ball. After planetary milling, the respective composite powders were each packed under argon into clean, 18cm long, alloy 110 copper billets (OD 9.53mm, ID 6.35mm), which were previously annealed at 650°C for 1 hour in an argon atmosphere. After packing, the billet ends are sealed with soft copper plugs and Wood's metal before removal from the inert atmosphere. The sealed billets were drawn to 0.984mm diameter wire using a hydraulic draw bench and a 10% reduction schedule [40]. All billets were reduced to the final wire diameter with no intermediate heat treatments.

X-ray powder patterns of post-milled $MgB_2$ and $MgB_2/M(20)$ composite powders were obtained on a Rigaku miniflex powder diffractometer using Cu K$\alpha$ radiation. These data indicate that there was no detectable degree of contamination from the milling media by XRD (~0.5% by volume sensitivity). However, trace amounts of Mg metal and oxides of magnesium and boron were detected by XRD. These are common contaminants in commercially available $MgB_2$.

Figure 1 shows polished transverse cross-sections and composite micrographs of the as-drawn wires. In general, all wires display good filament uniformity with an approximate filament diameter of 700μm. Figures 1(b) through (d) show the filament cross-section and micrograph of the $MgB_2/Ag(20)$, $MgB_2/In(20)$, and $MgB_2/Pb(20)$ wires, respectively. These filaments consist of a distribution of $MgB_2$ particles, with a maximum diameter of approximately 10μm, in intimate contact with the added metal. The planetary milling process has significantly reduced the average domain size of the starting In and Pb powders (<44μm and <74μm, respectively), and in all three samples has resulted in a nearly homogeneous distribution of material at the filament diameter length scale. The MgB$_2$/Ga(20) wire cross-section, shown in Figure 1(e), is unique in that the distribution of Ga metal within the filament is not uniform. The light regions in the Figure represent domains of high Ga content. In addition, there is a region at the interface between the Cu sheath and the MgB$_2$/Ga(20) filament in which there is significant alloying of the Cu and Ga. This is shown in Figure 2 where it is seen that there is an approximately 10μm wide interfacial region between the Cu and the filament core. Compositional analysis of this region indicates that it consists of approximately 63% atomic Ga. Figure 2 also shows a higher concentration of Ga metal within approximately 20μm of the sheath/filament interface. This increased concentration of Ga may result from the compressive forces experienced during the drawing of the wire, which force molten Ga out of the initially homogeneous mixture.

The starting billet filament fill factor % (i.e. the percentage of the total billet cross-section that is filament) for all samples is 44%. We find that this percentage is preserved during the wire drawing process for all samples [41], with a total measured variation in the filament fill factor % of approximately 5%. In contrast, we find that the filament packing fraction % (i.e. the percentage of actual mass of material in the filament to the theoretical limit) decreased from approximately 70 to 80% in the starting billet to approximately 60% for all samples when the billet was drawn into a wire [42]. Exceptions to this are the MgB$_2$/Pb(20) wire which decreased from 77% in the billet to 73% in the as-drawn wire, and the MgB$_2$ PIT wire which decreased dramatically from 68% in the billet to only 33% of the theoretical limit in the as-drawn wire.

Transport critical current I$_C$ measurements were determined using the 1µV/cm electric field criterion at temperatures ranging from 4.2K to 40K in self-field. Critical current densities J$_C$ were calculated using the average measured cross-sectional area of the filament. If possible, n-values were calculated from the electric field versus current data at electric fields in excess of 1µV/cm and less than 10µV/cm [43, 44, 45]. Unless indicated otherwise, the superconducting transport properties of these wires, at temperatures greater than 4.2K in self-field, were determined using a hairpin probe design. One disadvantage associated with this measurement probe is that it is necessary to bend the wire over a forming mandrel prior to the measurement. The result is that all transport I$_C$ measurements have been measured under approximately 2% bending strain, which results from bending a 0.984mm diameter wire over a 50mm diameter mandrel prior to mounting the sample on the probe [46]. For low transport currents ( i.e. < 10A ), we observe an approximately 20% reduction in I$_C$ determined using the hairpin probe versus a straight sample I$_C$ probe for all MgB$_2$/M(20) wires in this study. However, we find a 2 to 3 time reduction in the I$_C$ of the MgB$_2$ PIT wire when measured using the hairpin probe versus a straight sample I$_C$ probe, indicating the inherently low strain tolerance of as-drawn *ex situ* MgB$_2$ PIT wire. The increase in strain tolerance of *ex situ*

MgB$_2$ PIT wires with normal metals added to the core filament has been observed previously [35].

Electric field versus DC transport current measurements for MgB$_2$ and MgB$_2$/M(20) PIT wires at 10K in self-field are shown in Figure 3. The measured I$_C$, J$_C$, and n-value for these wires are listed in Table I. The MgB$_2$ PIT wire, with a J$_C$ of 0.85 kA/cm$^2$, is significantly lower than literature values which can be in excess of 100 kA/cm$^2$ at 10K. However, this J$_C$ is not unreasonable given the as-drawn (i.e. unannealed) nature of the wire and the unusually low filament packing fraction %. While the addition of 20% by volume Ag to the filament does little to enhance the transport properties of the wire, the addition of Pb, In, or Ga metal to the MgB$_2$ filament results in a significant increase in I$_C$ of the wire. In particular, the addition of 20% by volume Ga metal to the core filament results in a nearly 50 fold increase in the I$_C$ of the wire at 10K. This is a remarkable result given that the MgB$_2$/Ga(20) wire is the *least* homogeneous of all the wire samples. The MgB$_2$/Ag(20), MgB$_2$/Pb(20), and MgB$_2$/In(20) wires possess nearly identical microstructures (Figure 1(b)-(d)), yet show dramatic increases in both I$_C$ and n-value with the addition of Ag, Pb, and In, respectively.

The temperature dependent transport properties in self-field of MgB$_2$ and MgB$_2$/M(20) PIT wires are shown in Figure 4. These data were collected from several 25cm sections of wire selected from a central 5m section of wire. Because the temperature of the measurement probe is not under active control, data from at least four different wire sections at several different temperatures are plotted in the Figure. The low variance observed in the temperature dependence of the measured $I_C$ in these 25cm wire sections indicates a high degree of filament uniformity over the central 5m of wire. Figure 4(a) shows the $J_C$ and $I_C$ of all wires fabricated in this work. From the Figure it is seen that the $J_C$ of the $MgB_2/Ga(20)$ wire is approximately 50 times that of the $MgB_2/Ag(20)$ wire at temperatures up to approximately 25K. Figure 4(b) shows the n-value for all samples tested at temperatures greater than 10K. Again, the $MgB_2/Ga(20)$ shows the best performance with measured n-values in excess of 40 at 15K, and in excess of 20 at 20K in self-field. In general, the $MgB_2/M(20)$ series wires, all with approximately the same filament packing fraction %, display a clear trend of increasing $I_C$ and n-value with the addition of 20% by volume Ag, Pb, In, and Ga, respectively.

### III. Data Analysis

Our results indicate a clear enhancement in the current carrying properties of *ex situ* $MgB_2$-based PIT wires with the addition certain normal metals to the $MgB_2$ filament. Other groups have also observed similar increases in the $I_C$ of $MgB_2$-based PIT wires and have suggested several possible reasons for such an enhancement [33, 34, 36]. These include an improvement in the $MgB_2$ "grain linkage" resulting from an increased Josephson current between the grains or a proximity effect within the normal metal, an increased densification of the core filament, and an increase of the intra-filament thermal stability.

We analyze our results based on the schematic model of an ideal $MgB_2$/Metal composite shown in Figure 5. In this model, $MgB_2$ particles are embedded within a ductile metal matrix in which individual $MgB_2$ particles are in contact with each other at well-defined contact points, and the ductile metal is localized within the interstitial regions of the composite. Although there are numerous direct $MgB_2$/$MgB_2$ contacts in the composite, the total area of these contacts is small relative to the $MgB_2$/Metal/$MgB_2$ contact area. Thus, we propose that the transport $J_C$ of these composites may be described using the formalism developed to describe the $J_C$ of conventional Superconductor/Normal/Superconductor (*S/N/S*) junctions. Essentially, we treat this ideal composite as a conglomeration of *S/N/S* junctions.

We assume that all *S/N/S* junctions in the $MgB_2$/M(20) composite wires are in the clean limit ( $l >> \xi$ ), where *l* is inelastic mean free path of the electrons and $\xi$ is the coherence length in *N*. In this limit, Hsiang and Finnemore [47] showed that the critical current density of a thick *S/N/S* junction in zero applied magnetic field approximately follows the relation,

$$J_C = J_0 \, (1-T/T_C)^2 \, \exp(-K_N \, d) \,, \tag{1}$$

where $T_C$ is the superconducting critical temperature of *S*, *d* is the thickness of the normal metal layer *N*, $J_0$ is the maximum critical current density of the *S/N/S* junction at zero temperature, and $K_N^{-1}$ is the proximity effect decay length in *N*. In the limiting case where the electron-electron interactions in *N* are negligible, this decay length can be written as [48, 49],

$$K_N^{-1} = \hbar \, v_F/(2\pi \, k_B T) \qquad (2)$$

where $v_F$ is the Fermi velocity in *N*, and $k_B$ is Boltzmann's constant. Equation (2) is a simplified result, which is obtained when the temperature T is much greater than the superconducting critical temperature of the normal metal $T_{CN}$. As pointed out by Deutscher and de Gennes [49], when T approaches $T_{CN}$ in *S/N* junctions with non-zero electron-electron interactions in *N*, $K_N^{-1}$ diverges and equation (2) underestimates the extent of the proximity effect in *N*. In this work, the majority of the $J_C$ measurements have been obtained at temperatures well in excess of $3T_{CN}$ of the normal metal addition and thus, we neglect the divergence of $K_N^{-1}$ as T → $T_{CN}$ in the fitting of these $J_C$ data.

Figure 6 shows the results of fitting the $MgB_2$ PIT and $MgB_2$/M(20) $J_C$ data to equation (1) using literature values for the Fermi velocities of the metals [50, 51]. The best fit parameters are tabulated in Table II, along with specific relevant parameters evaluated at 20K. The measured $J_C$ of all wires in this study fit the model remarkably well using only three fitting parameters; *d*, $T_C$, and $J_0$. Referring to Equation (1) it is seen that the functional form of the critical current density in *S/N/S* junctions is completely determined by the critical temperature of the superconductor, and the thickness and Fermi velocity of the normal metal. The $J_0$ term is simply a scaling factor.

The results of these model fits reveal certain characteristics of the supercurrent transport mechanism in these composites. In particular, the thickness of the normal metal layer *d* we interpret as the average distance between individual $MgB_2$ particles in the $MgB_2$/M(20) composite filament, which carries the majority of the transport supercurrent. From Table II, it is seen that *d* is on the order of 1500Å for all $MgB_2$/M(20) samples. It is not surprising that this distance is on the order of the proximity effect decay length in *N* (e.g. $K_N^{-1}$ at 20K in Table II), as the majority of the supercurrent in the composite likely flows through the large area $MgB_2$/Metal/$MgB_2$ contacts within $K_N^{-1}$ of the *S/N* interfaces.

Consistent with our interpretation of the normal metal layer thickness *d* obtained from the model fits, is the very short inter-particle distance obtained from the fit of the $MgB_2$ PIT $J_C$ data. In this case, an inter-particle distance of approximately 300Å suggests that thin *metallic* regions between the $MgB_2$ grains limit the supercurrent in unannealed *ex situ* $MgB_2$ PIT wire. Annealing *ex situ* $MgB_2$ PIT wire increases the inter-grain contact area, relieves strain in the material, and reorders the surface layer of the individual grains. This effectively shrinks the thickness of the $MgB_2$/$MgB_2$ grain boundary contacts and establishes a high current *S/S* interface. Note that the inter-grain contact distance obtained from the temperature dependent $I_C$ of *ex situ* $MgB_2$ PIT wire, and the metallic nature of the contact, is consistent with the $MgB_2$ grain boundary studies of Samanta, *et al*. [7].

These model fits also reveal that the $T_C$ for all wire samples is approximately 33K to 34K, which is considerably lower than the bulk $T_C$ of commercial $MgB_2$. We interpret this $T_C$ as the critical temperature of the $MgB_2$ at the particle *surface* adjacent to the *N* metal, and not the bulk value. This is consistent with mechanical milling studies of $MgB_2$ which indicate that the $T_C$ of this material is lowered by the surface disorder and strain produced during the milling process [52]. All composites in this study were prepared using a planetary milling procedure and thus the $MgB_2$ surface $T_C$ is most likely lower than the bulk value.

As mentioned previously, the $J_0$ term in equation (1) represents the maximum $J_C$ of an *S/N/S* junction at zero temperature. In conventional S/N/S junctions this term is typically less than $10^7$ A/cm$^2$ in thin films [47, 53]. In our $MgB_2$ and $MgB_2$/M(20) composite wires, however, $J_0$ spans nearly two orders of magnitude and includes the macroscopic composite structure of the total wire cross section as well as the microscopic limits related to the maximum zero temperature $J_C$ of the *N* layer.

### IV. Discussion

Although the transport properties of the $MgB_2$ and $MgB_2$/M(20) PIT wires are well described by the *S/N/S* model described in the previous section, there is a question as to why there is a nearly 50 fold increase in the $I_C$ of the $MgB_2/Ga(20)$ wire versus the $MgB_2/Ag(20)$ wire. In principle, there are only three parameters that may account for the extremely large differences in the $J_C$ of these wires; $J_0$, $d$, and $T_C$. The $T_C$ values obtained from the fits to these data are all approximately the same, thus the $(1-T_R)^2$ term in equation (1) cannot contribute to the large observed increases in $J_C$. Table II lists the value of the $(1-T_R)^2$ term at 20K for all wire samples. Also, Table II lists the values of the $Exp(-K_N d)$ term in equation (1) at 20K, using the Fermi velocity and thickness of the *N* layer for all wire samples. Here, the shorter inter-particle distance of the $MgB_2/Ga(20)$ composite relative to the $MgB_2/Ag(20)$ composite accounts for a small increase in the $J_C$ of the composite on the order of a factor of 2. The only remaining parameter that can account for the increased $J_C$ of $MgB_2/M(20)$ is the $J_0$ term. This factor depends on the total surface area of the *S/N* contacts in the composite, which is related to the filament density, and the microscopic properties of the *N* layer which limit the supercurrent flow.

### A. $J_C$ Enhancement through Filament Densification

It is well known that critical current densities in granular superconducting wires and tapes may be increased by increasing the density of the superconducting filament. Critical currents increase with filament densification in PIT wire and tape from an increased contact area between adjacent superconductor grains in the filament. Filament porosity in $MgB_2$ PIT wire is generally minimized using sheath materials with high toughness (i.e.

Fe or Ni) [12], intermediate and final heat treatments, compressive wire deformation procedures, or flat rolling the wire to a tape geometry. The low filament packing % (33%) of the $MgB_2$ PIT wire relative to all $MgB_2/M(20)$ wires accounts for the dramatically lower transport $J_C$ of this wire. The increase in $I_C$ of the $MgB_2/M(20)$ wires with Ag, Pb, In, and Ga, however, is not as easily assigned to a filament densification model because the filament packing fraction % of these wires is approximately 60%, with the exception of Pb at 73%. The lack of a correlation between filament density and $I_C$ in $MgB_2/M(20)$ wires eliminates a simple filament densification model as an explanation for the increase in supercurrent transport of these wires.

Another possible reason for the enhanced $I_C$ of $MgB_2/M(20)$ wires may be because of the relative hardness of the added metals. In our *S/N/S* composite model, $I_C$ will increase with increased *S/N* contact area. The most dramatic increases will occur with increased *S/N* contact areas that possess *N* layer widths on the order of $K_N^{-1}$. The hardness of a material is its resistance to a localized plastic deformation. In the fabrication of $MgB_2/M(20)$ PIT wire, the forces which deform the billet during the drawing process will cause the hard, brittle $MgB_2$ particles to deform the softer metal particles. This process forces the softer metal into the interstitial regions of the $MgB_2$ particle network in the filament [34], and may significantly increase the overall *S/N* contact area. The Brinell hardness of In, Ag and Pb is 8.83, 24.5, and 38.3, respectively [54]. Thus, although it is difficult to quantify the increase in *S/N* surface area in the $MgB_2/M(20)$ composite, we would expect to see an increase in $I_C$ in with the softer metals. The fact that the $MgB_2/Pb(20)$ wire possesses an $I_C$ nearly 20 times that of the $MgB_2/Ag(20)$ wire is inconsistent with this model. However, the very high $I_C$ of the $MgB_2/Ga(20)$ wire may in part be the result of the increased *S/N* contacts in the filament, as the Ga is clearly in the liquid state during the wire deformation process.

## *B. $J_C$ Enhancement through the Superconducting Proximity Effect*

It appears likely that slight differences in the $MgB_2/M(20)$ composite morphology may account for changes in $J_C$ on the order of a factor of two, yet this increase remains well below the observed enhancement of the transport properties of $MgB_2/M(20)$ wires. We propose that the dramatically increased $J_C$ in these wires may be due to the unique properties of the specific metal addition that is in intimate contact with the superconducting $MgB_2$ particles in the filament.

It is well known that when a superconductor is placed in contact with a normal metal, the Cooper pair amplitudes in the superconductor do not vanish abruptly at the *S/N* interface, but extend a significant distance into *N*. This phenomenon, characterized by the induction of superconducting properties in the otherwise normal metal within $K_N^{-1}$ of the *S/N* interface, is known as the superconducting proximity effect [49, 55, 56, 57, 58]. Near the *S/N* interface, the normal metal exhibits superconducting properties such as a reduction in the electronic density of states at the Fermi surface [59, 60, 61], and a significant Meissner effect [62, 63, 64, 65, 66, 67, 68]. In addition, *S/N/S* junctions are known to support supercurrents over distances much greater than $K_N^{-1}$ at temperatures below $T_C$ of the *S* layer [69, 47, 53].

Here we suggest that the observed increase in $J_C$ of $MgB_2$/M(20) PIT wires results from the magnitude of the proximity effect induced superconducting gap Δ in *N*, which is specific to each metal addition. In this case, the *S/N/S* model junction discussed in Section III becomes an *S/S'/S* junction where the *S'* layer is weakly superconducting [70] and $J_0$ is limited by the magnitude of the minimum Δ in the *S'* layer. Though a detailed calculation of the magnitude of Δ in $MgB_2$/M(20) composites is an intractable problem, a reasonable first approximation may be obtained by solving for Δ in one-dimensional $MgB_2$/Metal model junctions.

Extending the work of de Gennes [48, 70], McMillan showed that the magnitude of Δ in *N* is a function of both the Cooper pair amplitude and the magnitude of the electron-electron interaction in *N* [56]. With a first approximation that Δ(x) is constant in *S* and zero in *N*, the spatial dependence of the superconducting gap Δ(x) in *S/N* junctions is determined self-consistently from,

$$\Delta(x) = \lambda^*(x) F(x) , \qquad (3)$$

with [71],

$$\lambda^* = (\lambda - \mu^*)/(1+\lambda) , \qquad (4)$$

where F(x) is the Cooper pair amplitude, λ is the electron-phonon coupling constant [72] and μ* is the magnitude of the screened Coulomb repulsion [73]. This general method of solving for Δ(x) in *S/N* junctions has been extended recently to include junctions with different Fermi velocities and electron mean free paths [58]. In our work, we use numerical methods developed in the study of proximity effects in *S/N/N'* junctions to calculate the magnitude of Δ(x) in model $MgB_2$/Metal junctions at finite temperature. We also include the effects of finite electron mean free paths and differing Fermi velocities in the *S* and *N* layers [74].

The Cooper pair amplitude F(x), normalized to its value deep within the bulk superconductor $F_{Bulk}$, for one dimensional $MgB_2$/Metal junctions is shown in Figure 7(a). These data are calculated at 20K using the parameters listed in Table III, the $v_F$ of $MgB_2$ listed in Table II, and an average Δ ~ 6.7 meV at 20K for $MgB_2$ [75]. In general, F(x) decreases with increasing distance from the *S/N* interface and this decrease is exponential with x for distances >> $K_N^{-1}$ [49]. These calculations show that the Cooper pair amplitude in *S/N* junctions is severely attenuated by short electron mean free paths in the *N* layer [76]. In the Figure, it is seen that F(x) extends several thousand Å into the *N* layer in the $MgB_2$/Ag and $MgB_2$/Ga model junctions because of the very long mean free paths of the electrons in Ag and Ga at 20K. In contrast, the high resistivity Pb layer in the $MgB_2$/Pb model junction has a much shorter mean free path, and thus the Cooper pairs do not extend as deeply into *N*.

Using the calculated Cooper pair amplitudes and the electron-electron coupling specific to each normal metal, the spatial dependence of the induced superconducting gap in *N* is obtained using equation (3). These data are shown in Figure 8(b), where it is clearly seen that $\Delta(x)$ in *N* scales linearly with $\lambda^*$. Normal metals with very low $\lambda^*$ values, do not exhibit strong induced superconductivity even though there may be a large Cooper pair amplitude in *N*. Here we make a distinction between a large amplitude of correlated electron pairs in *N*, and a large $\Delta$. The $MgB_2$/Ag model junction in Figure 7 shows this distinction clearly, where it is seen that even though F(x) is very large in the Ag layer, the extremely low $\lambda^*$ value specific to Ag metal results in a very small induced $\Delta$. On the other hand, the $MgB_2$/Ga model junction has both a large Cooper pair amplitude, and a large $\lambda^*$ in *N*. Thus, the Ga layer in the $MgB_2$/Ga model junction displays a large induced $\Delta$.

From these model calculations, it is evident that there is a correlation between the physical properties of the *N* layer material and the magnitude of the induced $\Delta$ in *N*, which we propose is responsible for the enhanced $J_C$ observed in $MgB_2$/M(20) PIT wires. In Figure 8 we plot the measured $J_C$ of each $MgB_2$/M(20) wire at 20K vs. the magnitude of the induced $\Delta$ from Figure 7(b) at 500Å. We approximate the *N* layer thickness in these three dimensional composites with a value of approximately 500Å in the one dimensional model, but also choose this value as it results in a very good correlation to these experimental data. This correlation is perhaps better than we should expect to obtain given the numerous simplifications of our model, yet we believe it provides strong evidence that the $J_C$ of $MgB_2/M(20)$ composite wires is largely determined by a proximity effect induced $\Delta$ in the normal metal component of the composite.

Critical supporting evidence of the existence of a proximity effect limited $J_C$ in $MgB_2/M(20)$ composite wires can be obtained from zero-field cooled (ZFC) DC magnetic susceptibility measurements. We follow the development of Thompson, *et al.* [77] where "non-ideal" granular superconductors were studied with DC magnetic methods. In granular superconductors, the intra-grain critical currents may be well in excess of the inter-grain critical currents. This is the case for example, in the HTS ceramics where weak link contacts at the grain boundaries are known to severely reduce the transport $I_C$ of HTS PIT tape [78]. The magnitude and temperature dependence of the intra- and inter-grain $I_C$ of granular superconductors may be determined using ZFC DC magnetic susceptibility measurements.

Figure 10 shows the ZFC DC magnetic susceptibility of all $MgB_2/M(20)$ composite wire in this study. These measurements were made on a Quantum Design SQUID magnetometer in an applied magnetic field of 10Oe. These data were obtained on 0.984mm diameter $MgB_2/M(20)$ wire samples approximately 5mm in length using the measured filament packing fraction %, and are plotted as the volume susceptibility (emu/cm$^3$). Thus, a direct comparison can be made between the as-measured susceptibility without including the geometric demagnetization factor for a cylinder [77]. As seen in Figure 9, the *onset* of a diamagnetic susceptibility in all samples occurs near 38K, and there is an increasing diamagnetic response of the MgB$_2$/M(20) composites with increasing λ* and $l$ of the *N* layer.  In particular, as the MgB$_2$/Ga(20) sample is warmed in an applied magnetic field, the diamagnetic susceptibility decreases slightly up to approximately 20K, then exhibits an abrupt loss of diamagnetism near 30K. This is in contrast to the MgB$_2$/Ag(20) composite in which the relatively small diamagnetic response shows no distinct transition with increasing temperature. The MgB$_2$/Pb(20) and MgB$_2$/In(20) composites, however, display a sudden loss of diamagnetism with warming at approximately 15K and 20K, respectively. We associate the sudden loss of diamagnetism in the MgB$_2$/Pb(20), MgB$_2$/In(20), and MgB$_2$/Ga(20) composite wires with the temperature at which the induced surface currents exceed the inter-grain I$_C$. In general, these data show there is a significant *increase* in the volume of *well-connected* (i.e. superconducting) contacts in the MgB$_2$/M(20) composites with high λ* normal metal additions with long electron mean free paths. This is a clear indication of the existence of a substantial proximity effect in these composite materials.

**Conclusions**

We have fabricated a series of *ex situ* copper stabilized $MgB_2$ PIT wires with different normal metals added to the core filament. We find that the transport critical current of these wires increases significantly with the addition of metals that possess both large electron-electron interactions and long electron mean free paths, and not due to the filament morphology. In particular, the temperature dependence and magnitude of the transport critical currents of $MgB_2$/M(20) wires is well described as an ensemble of *S/N/S* junctions in which the supercurrent is carried primarily by $MgB_2$/Metal/$MgB_2$ contacts with metal layer widths less than the de Gennes proximity effect decay length, $K_N^{-1}$. One-dimensional calculations of the magnitude of the proximity effect superconducting gap in $MgB_2$/Metal model junctions support this, and ZFC DC magnetization data clearly indicate an increased number of well-connected superconducting contacts in the $MgB_2$/M(20) composites which contain high $\lambda^*$ metals with long *l*. These results provide strong evidence that the supercurrent in $MgB_2$/M(20) composite wires is determined by inter-granular *S/N/S* critical currents, which are in effect determined by the magnitude of the proximity effect induced superconducting gap within the *N* layer.

The driving force for this work is to develop a means to establish superconducting inter-grain contacts in *ex situ* $MgB_2$ PIT wire without the use of heat. This would not only allow for the use of low cost copper as a stabilizing matrix for these conductors, but in principle should lower the fabrication costs with the elimination of a high temperature batch anneal. Further, the use of metal additions in these conductors has certain advantages, some of which have been documented in the literature and confirmed here. In particular, we have observed a significant increase in the strain tolerance of these wires as compared to *ex situ* $MgB_2$ PIT wire with no added metal. Also, the metal addition allows for the possibility of enhanced intra-filament stability due to the relatively high thermal conductivity of the added metal. This may be of particular importance in these conductors as it may be difficult to achieve the 1-5μm filament diameter typical of fully stabilized multifilamentary superconducting wire because of the relatively large $MgB_2$ particle size. Another advantage of this approach is that it allows for the addition of artificial pinning centers to the *normal metal layer* at relatively low temperatures. This may significantly improve the transport properties of $MgB_2$/Metal composite wires in applied magnetic fields.

Technical $MgB_2$ conductors, in addition to being low cost, must have high transport currents in modest magnetic fields. In general, superconducting applications such as transformers, motors, and generators require critical currents of at least 100A in approximately 1 to 2 Tesla at the operating temperature of the device. We have limited data on the transport properties of these conductors in applied magnetic fields. Figure 10(a) shows the transport critical current of $MgB_2$/Ga(20) wire at 4.2K in applied magnetic fields up to 6T [79]. The $J_C$ of this wire decreases exponentially with increasing magnetic field as anticipated from the behavior of *S/N/S* junctions in applied magnetic fields [47]. The n-value of $MgB_2/Ga(20)$ wire as a function of applied magnetic field is shown in Figure 10(b), where it is seen to be in excess of 20 at 2T at 4.2K. Assuming a similar magnetic field response at 20K, we estimate that the $MgB_2/Ga(20)$ wire will have a transport $J_C$ in excess of $5 \times 10^3$ A/cm$^2$ in 1T magnetic field.

The measured transport critical currents in $MgB_2/M(20)$ wires are less than what is required at present for practical superconducting devices, however, there is considerable room for improvement in these wires. Generally, $I_C$ optimization based on the composite packing density, $MgB_2$/Metal volume %, $MgB_2$ particle size, and conductor geometry are relatively simple engineering improvements to these conductors. Other improvements can be achieved using additional characterization/fabrication methods. In particular, we find the $MgB_2/Ga(20)$ wire has a relatively inhomogeneous cross-section, yet it possesses the highest transport current. This suggests there are regions within the wire that have *very* high critical currents. Magneto-optical studies performed on these conductors should help to elucidate where the supercurrent is flowing in this inhomogeneous composite. These results, in turn, should lead to significant improvements in the $I_C$ of these conductors. Also, the insertion of thin barrier layers, similar to those used in the copper stabilized *in situ* $MgB_2$ multifilamentary wire fabricated by Sumption *et al.* [17], will significantly reduce the alloy formation at the copper/filament interface in $MgB_2/Ga(20)$ wire and may also lead to improved transport currents. In addition, the relatively low $T_C$ at the $MgB_2$ particle surface is also limiting the supercurrent in these composites. A low-temperature post-mill anneal of the composite before packing into the billet may minimize this effect. Finally, because the superconducting proximity effect is known to be severely attenuated by the presence of insulating contaminants at the *S/N* interface, high quality $MgB_2$ powder is essential for the fabrication of high current $MgB_2$/Metal PIT wire. Thus, significant improvements in the transport properties of $MgB_2$/Metal PIT wire are expected with the use of high quality, high $T_C$ $MgB_2$ powder.

**Acknowledgements**

The author would like to thank Dr. David A. Tyvoll, Professor William A. Little, Professor James P. Collman, and Dr. Todd Eberspacher for many valuable discussions, technical assistance, and comments. In addition, the author gratefully acknowledges Dr. Ken A. Marken (Oxford Superconductor Technology) for providing transport data as a function of applied magnetic field, Dr. Reza Loloee (Michigan State University, Center for Fundamental Materials Research) for valuable assistance in obtaining the DC susceptibility data, and Mike Tomsic (HyperTech Research), Dr. M.D Sumption and Dr. E.W. Collings (Ohio State University, Laboratory for Applied Superconductivity and Magnetism) for many useful discussions regarding superconducting wire. Partial financial support provided by the California Energy Commission's Public Interest Energy Research group's Energy Innovations Small Grant Program.**References**

30financial support provided by the California Energy Commission's Public Interest Energy Research group's Energy Innovations Small Grant Program.

**References**

[1] J. Nagamatsu, N. Nakagawa, T. Muranaka, Y. Zenitani, and J. Akimitsu, Nature **410** (2001) 63.

[2] X.L. Wang, Q.W. Yao, J. Horvat, M.J. Qin, and S.X. Dou, Supercond. Sci. Technol. **17** (2004) L21.

[3] H. Kumakura, H. Kitaguchi, A. Matsumoto, and H. Hatakeyama, Appl. Phys. Lett. **84**, 3669 (2004).

[4] T. P. Sheahen and A.M. Wolsky, in *Introduction to High Temperature Superconductivity,* Ch. 16, p. 317, (Plenum Press, New York) 1994.

[5] D.C. Larbalestier, L.D. Cooley, M.O. Rikel, A.A. Polyanskii, J. Jiang, S. Patnaik, X.Y. Cal, D.M. Feldmann, A. Gurevich, A.A. Squitierl, M.T. Naus, C.B. Eom, E.E. Hellstrom, R.J. Cava, K.A. Regan, N. Rogado, M.A. Hayward, T. He, J.S. Slusky, P. Khallfah, K. Inumaru, and M. Haas., Nature **410**, 186 (2001).

[6] D.K. Finnemore, J.E. Ostenson, S.L. Bud'ko, G. Lapertot, and P.C. Canfield, *Phys. Rev. Lett.* **86**, 2420 (2001)

[7] S.B. Samanta, H. Narayan, A. Gupta, A.V. Narlikar, T. Muranaka, and J. Akimitsu, Phys. Rev. **B65**, 092510-1 (2002).

[8] S. Soltanian, X.L. Wang, I. Kusevic, E. Babic, A.H. Li, M.J. Qin, J. Horvat, H.K. Liu, E. Lee, E.W. Collings, M.D. Sumption, and S.X. Dou, Physica C 361, 84 (2001).

[9] X.L. Wang, S. Soltanian, J. Horvat, A.H. Liu, M.J. Qin, H.K. Liu and S.X. Dou, Physica C 361, 149 (2001).

Krusin-Elbaum, and F. Holtzberg, in *Magnetic Susceptibility of Superconductors and Other Spin Systems*, ed. R.A. Hein *et al.*, Plenum Press (1991).

[78] T. P. Sheahen, in *Introduction to High Temperature Superconductivity,* Ch. 13, p. 243, (Plenum Press, New York) 1994.

[79] Data provided by Dr. K.A. Marken, Oxford Superconducting Technology, Carteret, NJ.

# Figure Captions

**Figure 1.** SEM microstructures of $MgB_2$ and $MgB_2/M(20)$ wires. The left column shows the transverse cross-section of the as-drawn wires and the right column shows the average microstructure of the core filament.

**Figure 2.** SEM micrograph of the Copper/$[MgB_2/Ga(20)]$ interface. The ~10μm wide region at the interface consists of a CuGa alloy containing approximately 63% atomic Ga.

**Figure 3.** Electric field versus current measurements for $MgB_2$ and $MgB_2/M(20)$ PIT wires at 10K in self field. The slight negative bias of these data at high transport currents (>100A) results from an inductive pick-up in the voltage leads. The dashed line in the Figure indicates the 1μV/cm electric field criterion at which $I_C$ is determined. Note the $MgB_2/Ga(20)$ PIT wire tends to quench at transport currents in excess of 150A.

**Figure 4.** Temperature dependent transport properties of $MgB_2$ and $MgB_2/M(20)$ PIT wires in self-field. (a) The critical current density ($J_C$) and critical current ($I_C$) of all wires as determined using the 1μV/cm electric field criterion. (b) The n-value of all wires as determined from the electric field versus current data from 1μV/cm to 10μV/cm.

**Figure 5.** Schematic diagram of an ideal $MgB_2$/Metal composite used to interpret the transport properties of $MgB_2/M(20)$ PIT wire. In this model, the large $MgB_2$/Metal contact area determines the transport properties of the composite. Thus, these composites are essentially a conglomeration of *S/N/S* junctions.

**Figure 6.** Temperature dependent transport properties of $MgB_2$ and $MgB_2/M(20)$ PIT wires in self-field (symbols) and the best fit of these data using a *S/N/S* model in the clean

limit (solid lines). From the Figure it is easily seen that there is very good agreement between experiment and theory based on a simple *S/N/S* model describing the transport properties of these wires.

**Figure 7.** Proximity effect calculations for one-dimensional $MgB_2$/Metal junctions at 20K. (a) Cooper pair amplitude F(x) as a function of distance for each model junction normalized to the amplitude of the Cooper pairs deep within the bulk superconducting $MgB_2$. (b) Superconducting gap $\Delta(x)$ as a function of distance for each model junction. The magnitude of $\Delta(x)$ in *N* is proportional to the magnitude of the electron-electron interaction $\lambda^*$ in *N*.

**Figure 8.** Experimental transport properties of $MgB_2$/M(20) PIT wires versus the calculated magnitude of a proximity induced superconducting gap at 500Å in the normal metal layer. This correlation provides strong evidence that the transport properties of $MgB_2$/M(20) PIT wires are largely determined by the magnitude of the proximity effect in the *N* layer, which is specific to each metal addition.

**Figure 9.** Zero-field cooled DC magnetic susceptibility of $MgB_2$/M(20) PIT wire. These data show a clear increase in the well-connected diamagnetic volume of superconducting material in these composites, which contain metals with large electron-electron interactions and long electron mean free paths. These data provide clear evidence of a significant proximity effect in $MgB_2$/M(20) composite wires with high $\lambda^*$ metal additions.

**Figure 10.** Magnetic field dependent transport properties of $MgB_2$/Ga(20) PIT wire at 4.2K. (a) The critical current density ($J_C$) and critical current ($I_C$) as determined using the 1µV/cm electric field criterion. (b) The n-value as determined from the electric field versus current data. The applied magnetic field was oriented parallel to the wire axis for all measurements.

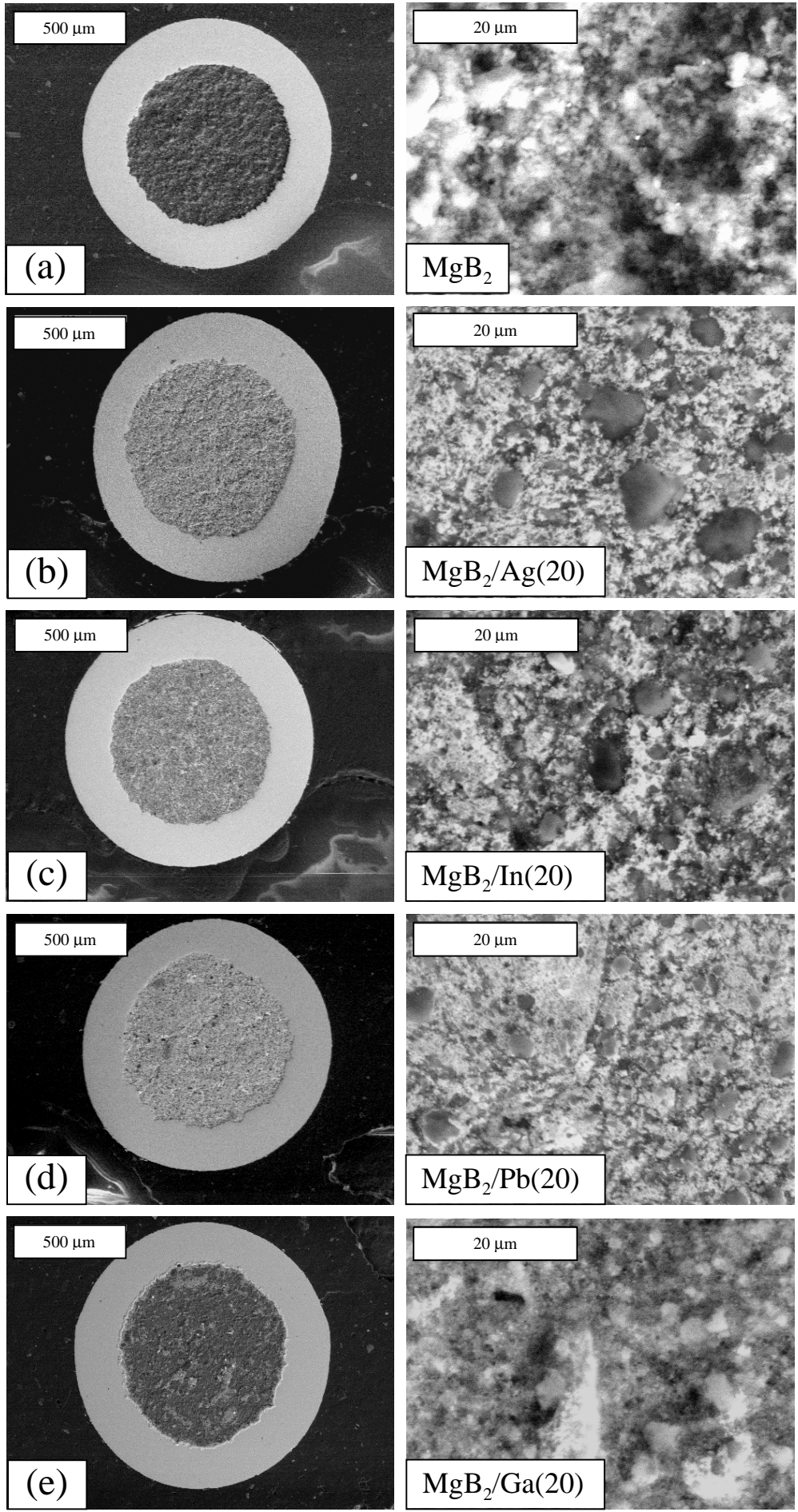

*Figure 1. Holcomb*

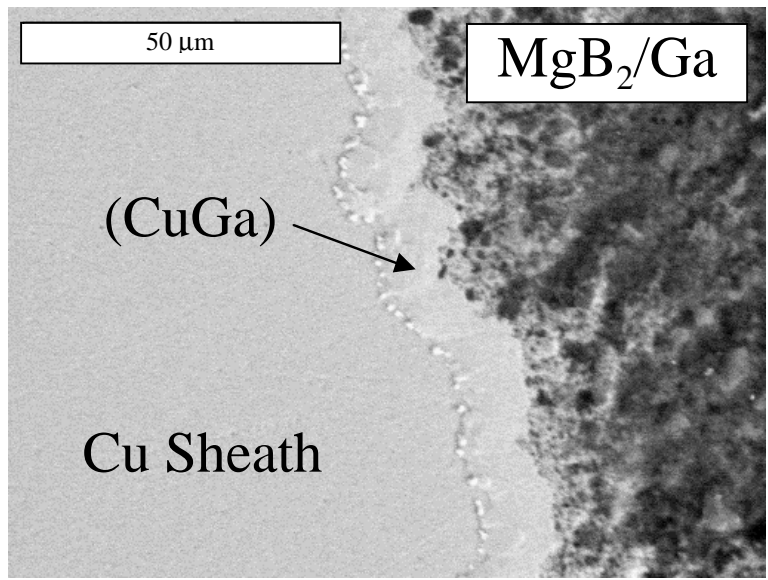

*Figure 2. Holcomb*

Figure 3. Holcomb

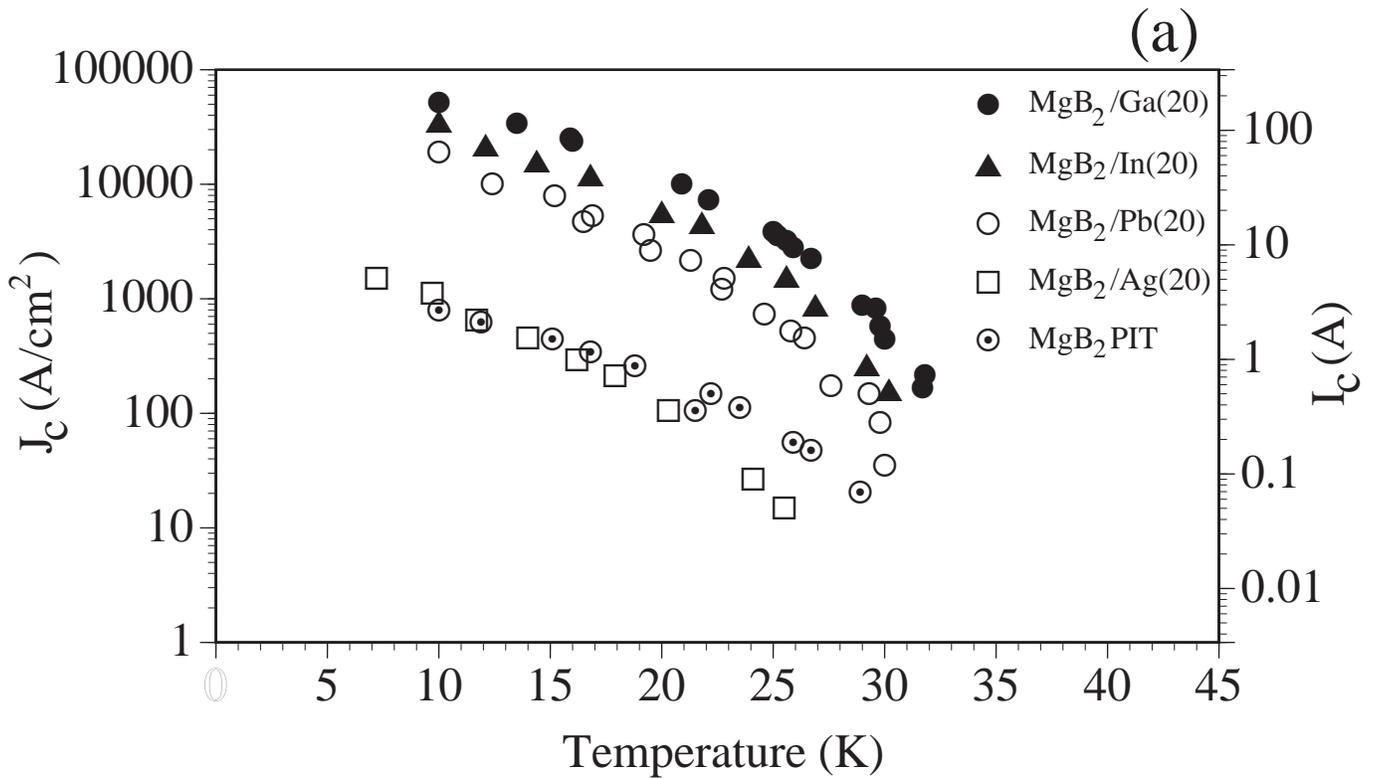

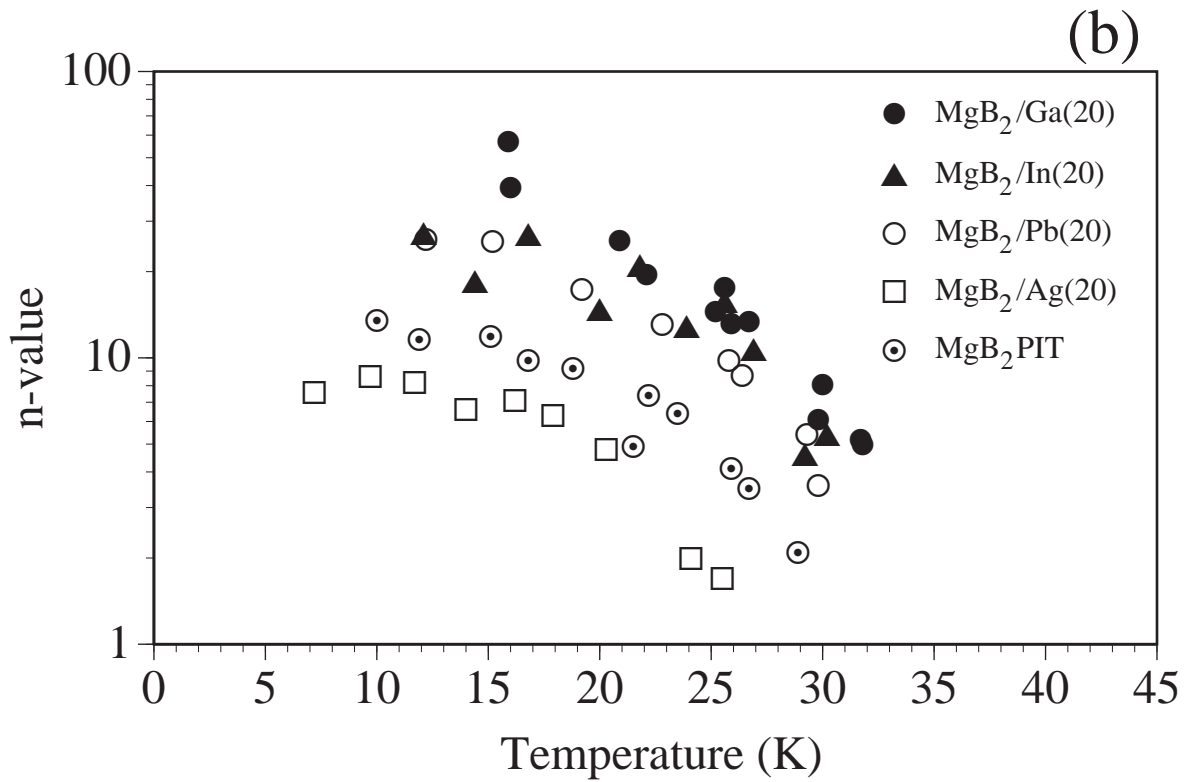

*Figure 4. Holcomb*

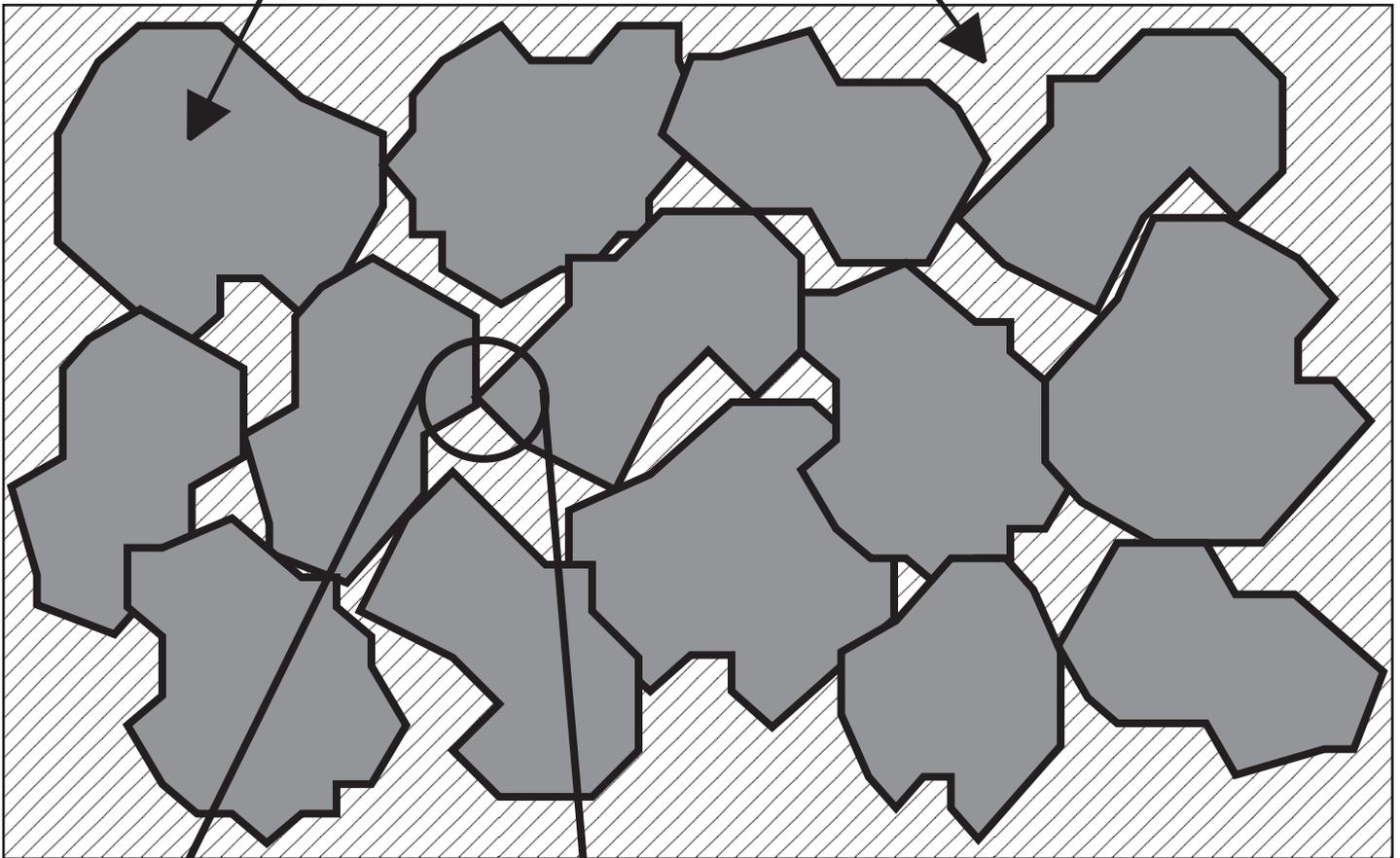
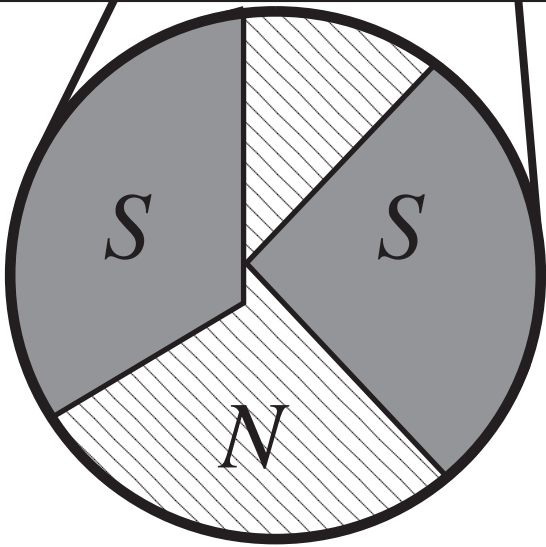
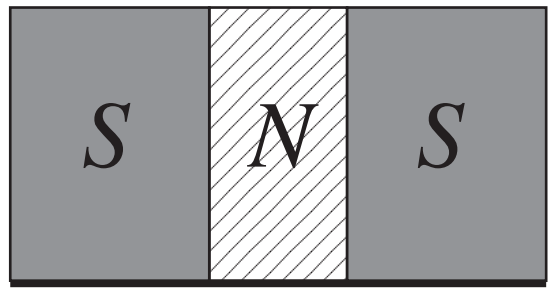

*Figure 5. Holcomb*

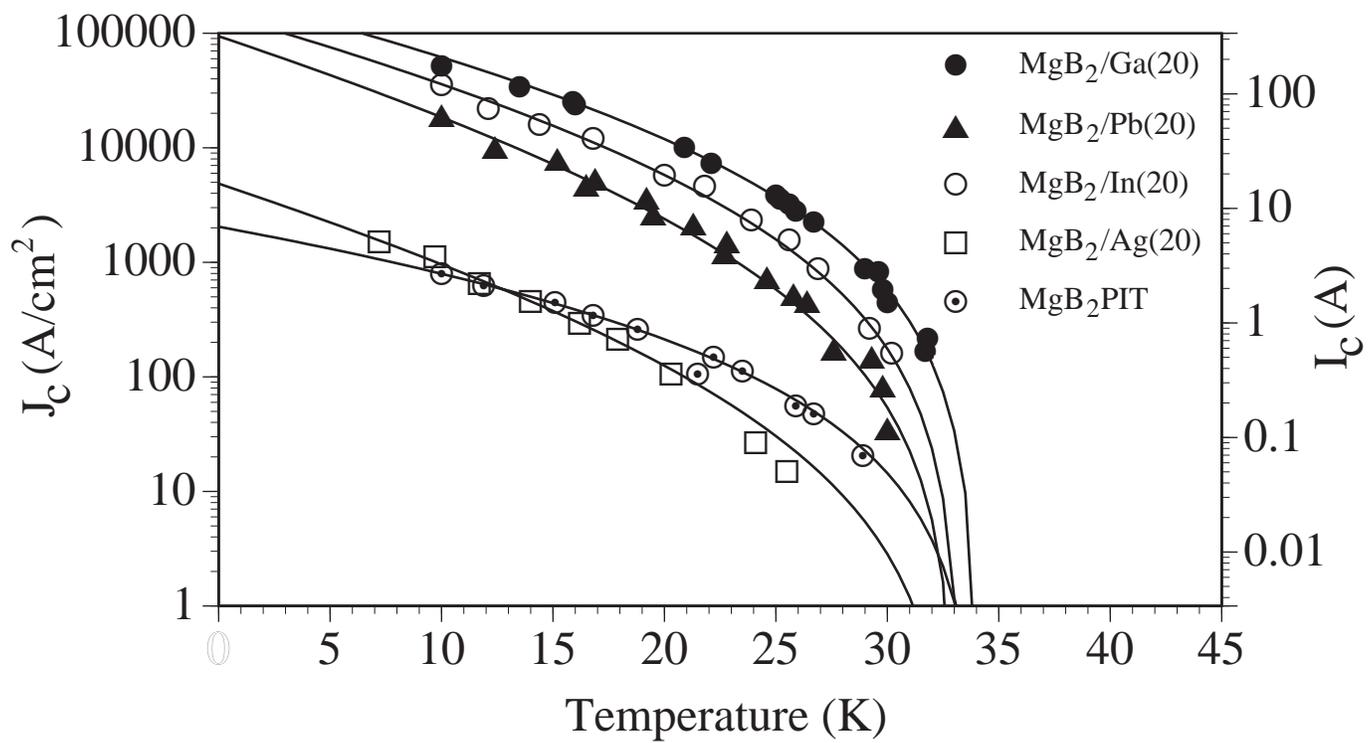

Figure 6. Holcomb

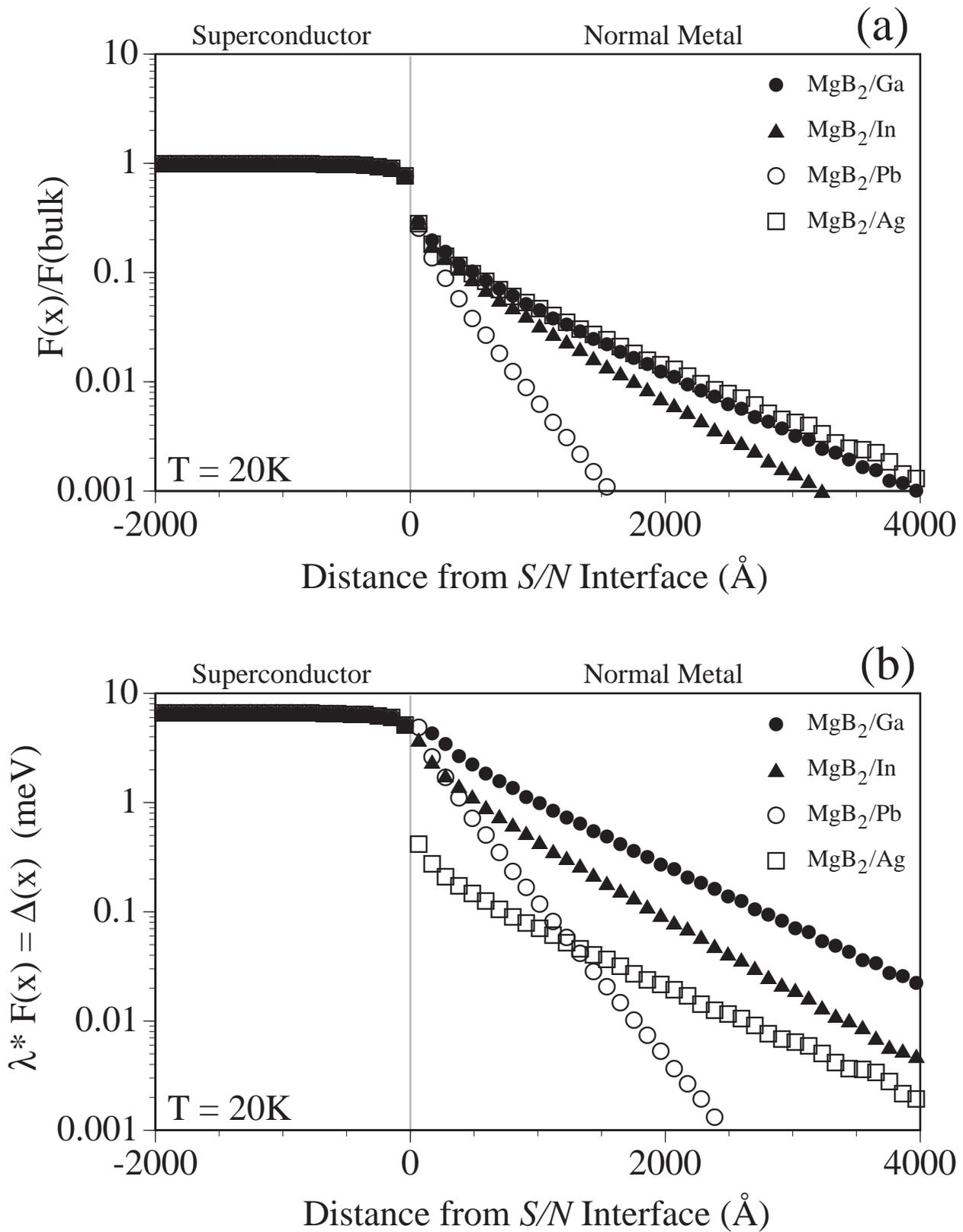

Figure 7. Holcomb

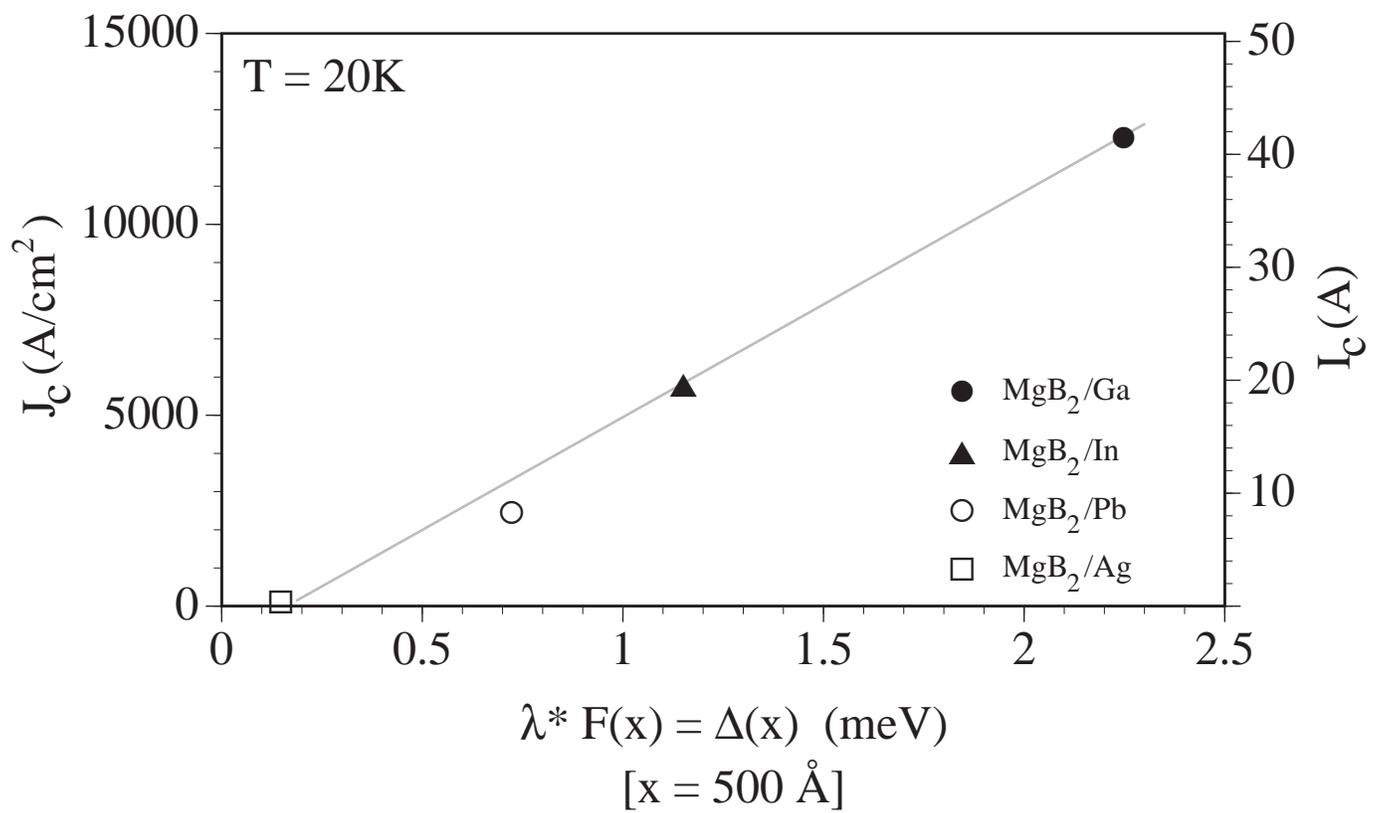

Figure 8. Holcomb

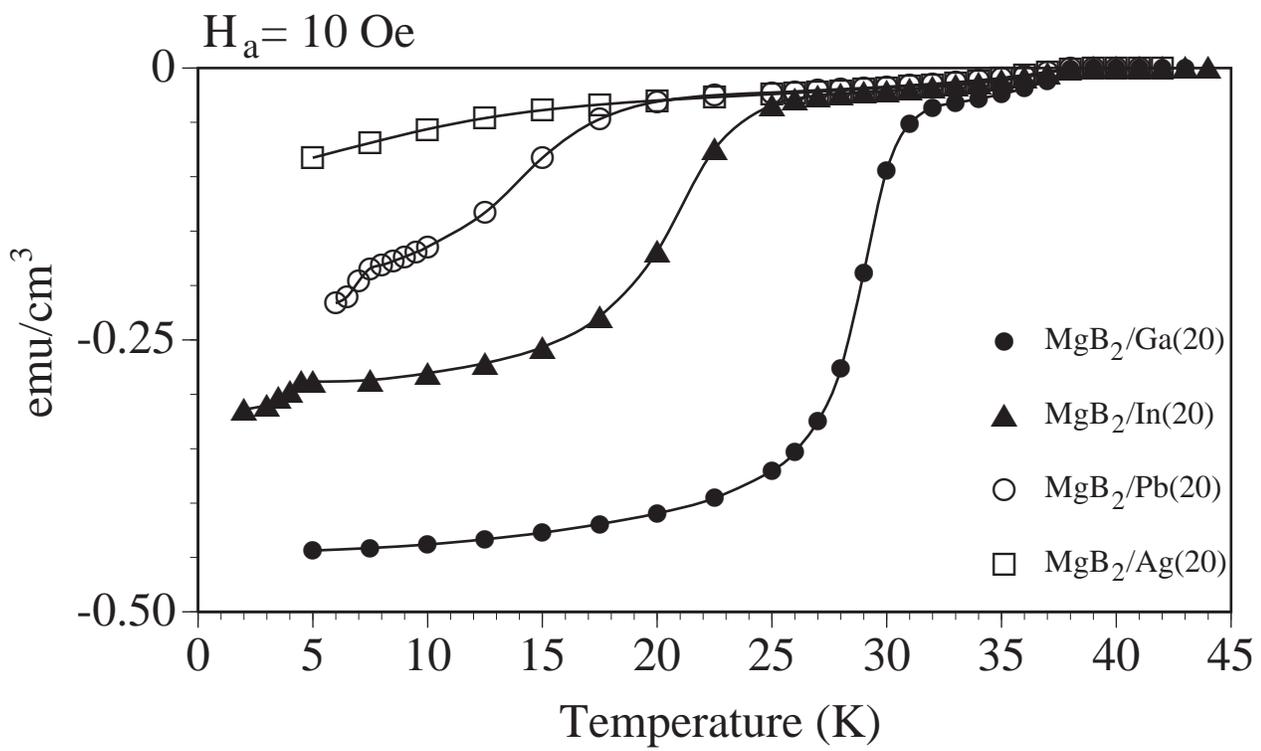

Figure 9. Holcomb

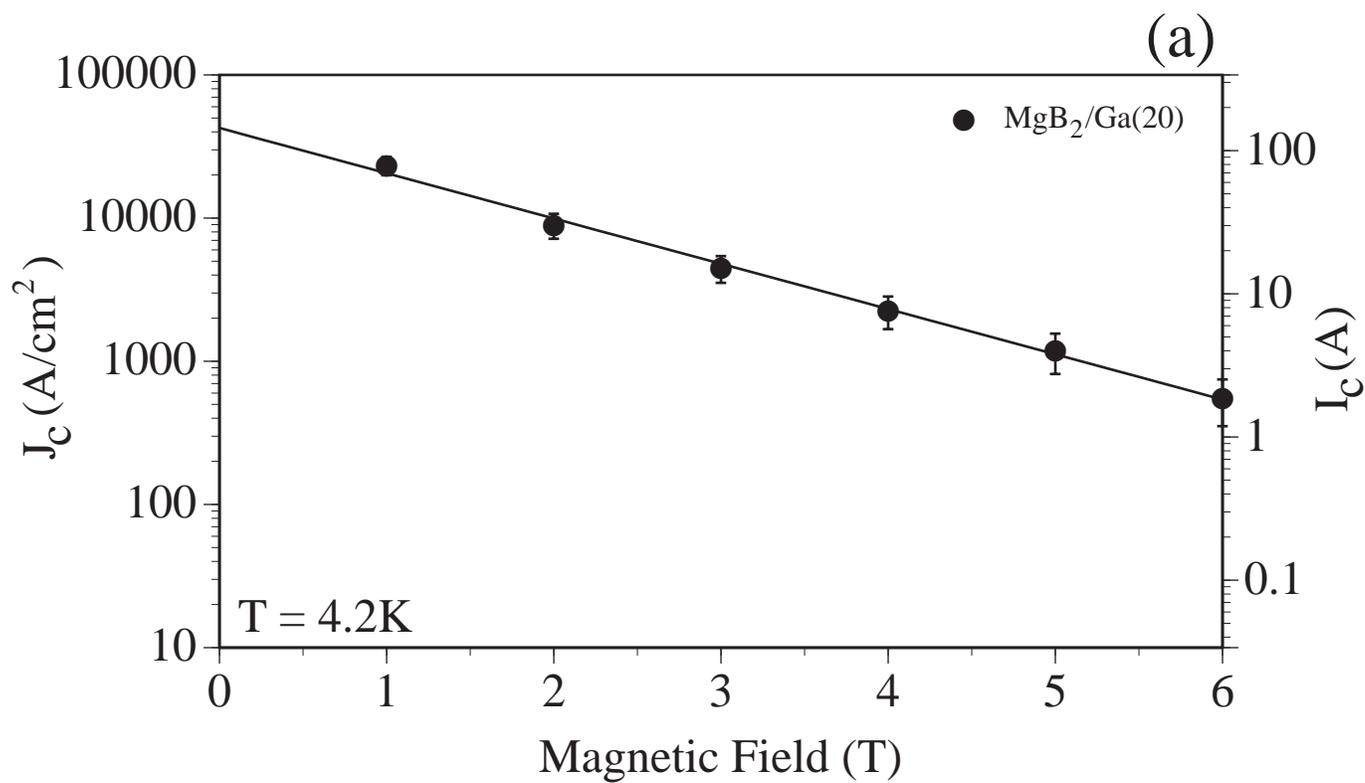
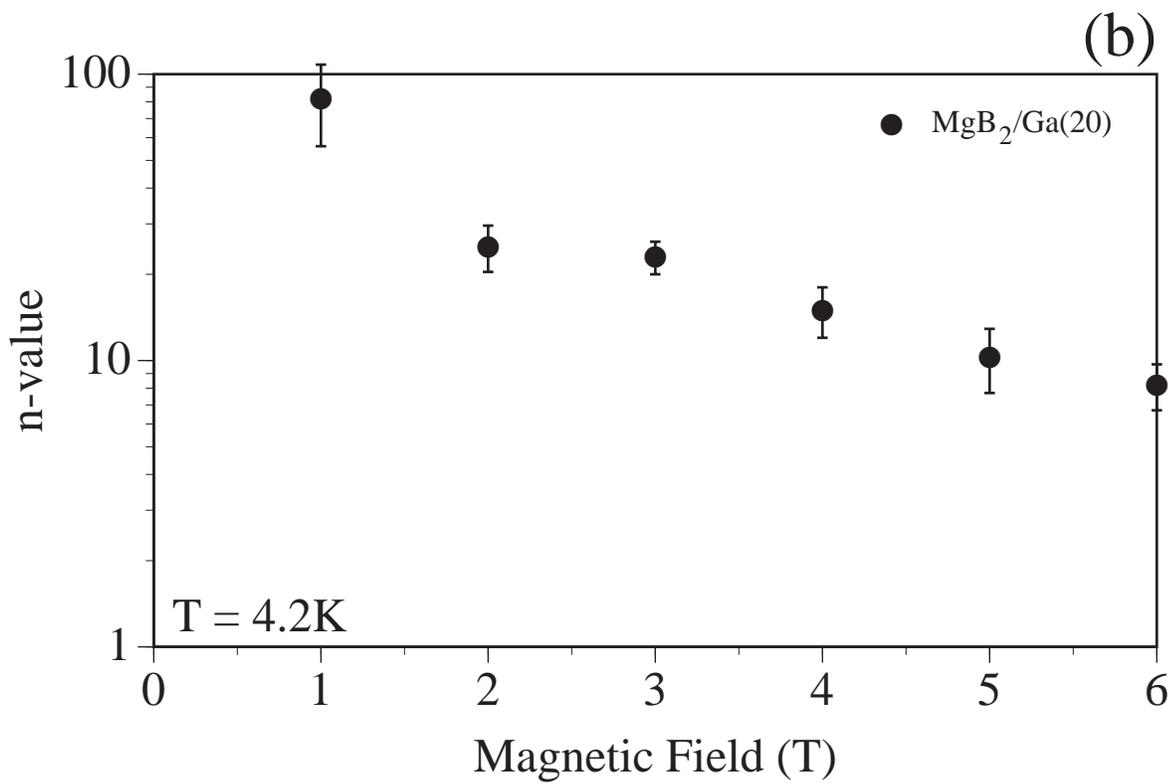

Figure10. Holcomb

| Sample | $I_C$ (A) | $J_C$ (kA/cm$^2$) | n-value |
|---|---|---|---|
| MgB$_2$ | 2.7 | 0.85 | 14 |
| MgB$_2$/Ag(20) | 3.8 | 1.11 | 9 |
| MgB$_2$/Pb(20) | 65 | 18.2 | 14 |
| MgB$_2$/In(20) | 119 | 36.4 | 53 |
| MgB$_2$/Ga(20) | 175 | 52.3 | -- |

**Table I**. Critical current, critical current densities and n-values of MgB$_2$ and MgB$_2$/M(20) wires at 10K in self-field.

| Sample | $J_0$ (kA/cm$^2$) | $d$ (Å) | $T_C$ (K) | $v_F$ ($10^8$ cm/s) | $K_N^{-1}$ (Å) at 20K | $\exp(-K_N^{-1}d)$ at 20K | $(1-T/T_C)^2$ at 20K |
|---|---|---|---|---|---|---|---|
| MgB$_2$/Ag(20) | 4.9±0.1 | 1566±50 | 33.1±1.1 | 1.39 | 845 | 0.157 | 0.156 |
| MgB$_2$/Pb(20) | 94.6±13.4 | 2043±270 | 33.1±0.8 | 1.83 | 1112 | 0.159 | 0.156 |
| MgB$_2$/In(20) | 147.0±4.6 | 1507±42 | 33.3±0.5 | 1.74 | 1058 | 0.241 | 0.160 |
| MgB$_2$/Ga(20) | 219.0±14.7 | 1336±125 | 34.1±0.4 | 1.92 | 1167 | 0.318 | 0.171 |
| MgB$_2$ PIT | 2.0±0.1 | 274±100 | 34.2±1.5 | 0.89 | 541 | 0.603 | 0.172 |

**Table II.** Parameters used in the fit of the experimental $J_C$ data.

| Material | $v_F$ ($10^8$ cm/s) | $\rho$ ($\mu\Omega$ cm) at 20K | $l$ (Å) at 20K | $\lambda$ | $\mu^*$ | $\lambda^*$ | $K_N^{-1}$ (Å) at 20K |
|---|---|---|---|---|---|---|---|
| Ag | 1.39 | 0.034 | 24,679 | 0.15 | 0.1 | 0.04 | 845 |
| Pb | 1.83 | 0.56 | 869 | 1.55 | 0.144 | 0.55 | 1112 |
| In | 1.74 | 0.16 | 3340 | 0.81 | 0.116 | 0.38 | 1058 |
| Ga | 1.92 | 0.09 | 4903 | 2.25 | 0.174 | 0.64 | 1167 |

**Table III.** Parameters Used in the Calculation of $\Delta(x)$ in $MgB_2$/M Junctions.